\documentclass[
    ,final            
  ]
  {aipproc}

\layoutstyle{8x11single}
\newcommand{\hI}{\hspace{1cm}}
\newcommand{\hVII}{\hspace{.7cm}}
\newcommand{\hV}{\hspace{.5cm}}
\newcommand{\mn}{\mu\nu}

\begin{document}

\title{Preheating the Universe from the Standard Model Higgs}


\author{Daniel G. Figueroa}{
  address={Instituto de F\'isica Te\'orica {\rm UAM/CSIC} and Departamento de F\'isica Te\'orica, Facultad\\
 de Ciencias, Universidad Aut\'onoma de Madrid, Cantoblanco, Madrid 28049, Spain.}
}



\begin{abstract}
We discuss Preheating after an inflationary stage driven by the Standard Model (SM) Higgs field non-minimally coupled to gravity. We find that Preheating is driven by a complex process in which perturbative and non-perturbative effects occur simultaneously. The Higgs field, initially an oscillating coherent condensate, produces non-perturbatively $W$ and $Z$ gauge fields. These decay very rapidly into fermions, thus preventing gauge bosons to accumulate and, consequently, blocking the usual parametric resonance. The energy transferred into the fermionic species is, nevertheless, not enough to reheat the Universe, and resonant effects are eventually developed. Soon after resonance becomes effective, also backreaction from the gauge bosons into the Higgs condensate becomes relevant. We have determined the time evolution of the energy distribution among the remnant Higgs condensate and the non-thermal distribution of the SM fermions and gauge fields, until the moment in which backreaction becomes important. Beyond backreaction our approximations break down and numerical simulations and theoretical considerations beyond this work are required, in order to study the evolution of the system until thermalization.  
\end{abstract}

\maketitle


\section{Introduction}

If the Universe went through an inflationary expansion in an early stage of its evolution, it must also have undergone a Reheating period afterwards, during which (almost) all the matter of the Universe was created. A simple consequence of Inflation is that the number density of any particle species exponentially dies away with the inflationary expansion. Therefore, at the end of inflation, the Universe has no particles at all. Only the homogeneous energy density responsible for inflation (usually the potential energy of a scalar field, the \textit{inflaton}) has survived. So, what?

The hot Big Bang (hBB) theory describes the early Universe as an expanding space filled up with relativistic species in thermal equilibrium. Thus, we are left on one hand, with an empty universe at the end of inflation and, on the other hand, with a universe full of relativistic particles according to the hBB. So, how can those two periods be matched? Obviously, the energy density responsible for inflation - the potential energy of the inflaton - had to get converted somehow into particles. The precise epoch in which the primordial inflationary energy was converted into (almost) all the particles of the Universe, is known as Reheating. 

How can particles be created out of the energy of one field (the inflaton)? The idea is that by coupling the inflaton to other fields of matter, then the potential energy of the inflaton gets converted into quanta of those fields. A specific scenario of Reheating will consist of a specific model of particle physics, specifying a particular form of the inflaton couplings to other matter fields. Depending on the model, there will be specific mechanisms of particle production taking place. For instance, the main mechanism of production might be \textit{perturbative decays}~\cite{Perturbative}, \textit{parametric resonance}~\cite{Preheating}, \textit{tachyonic instabilities}~\cite{Hybrid}, etc. Whatever the mechanism of particle production, the created particles interact among themselves and eventually reach a thermal equilibrium state with a common temperature. That is the \textit{Reheating Temperature}, $T_{\rm RH}$, which represents the energy scale at the end of Reheating and determines the first moment in which the description of the evolution of the Universe can finally be put within the hBB theory framework.

In practice, in order to analyze Reheating, we will study the dynamics of a system described by a potential
\begin{equation}
 V = V_{\rm inf}(\chi) + V_{\rm int}(\chi,\phi_a,\psi_b,A^c_\mu,...)\,,
\end{equation}
where $V_{\rm inf}$ is the inflationary potential and $V_{\rm int}$ represents the interaction terms between the inflaton $\chi$ and the rest of matter fields of the model: scalar fields ($\lbrace\phi_a\rbrace$), fermions ($\lbrace\psi_b\rbrace$), vector fields ($\lbrace A^c_\mu\rbrace$), etc. In the last few years, simplified models based only on scalar fields or, at most, including (non-gauge) fermionic species, have been the main target of study. For example, a simple model in which the inflaton $\chi$ is coupled to another scalar field $\phi$ and to some fermionic field $\psi$, can be described by
\begin{equation}\label{eq:potentials}
 V = V_{\rm inf} + V_{\rm int} = V_{\rm inf}(\chi) + g^2\chi^2\phi^2 + y^2\chi\bar\psi\psi\,,
\end{equation}
where $g^2$ and $y^2$ are dimensionless couplings. Unfortunately, even dealing only with two or three fields, the aspects of Reheating are extremely complex~\cite{Preheating},~\cite{FermionicPreheating}. Consequently, most phenomena of particle creation have been discovered in simplified scenarios like the one described by~(\ref{eq:potentials}), where the Standard Model (SM) particles or Dark Matter (DM) candidates are simply absent. Models incorporating a gauge principle in some sector of scalar fields (ignoring fermions), have also been considered. For instance, Hybrid Inflation models where the inflaton $\chi$ is a singlet of the SM, and the symmetry breaking field $\Phi$ coupled to the inflaton, is the SM Higgs doublet. Such models have been studied~\cite{BellidoTonyMarga} through lattice numerical integration of the classical equations of motion obtained from the Lagrangian
\begin{equation}
-\mathcal{L} = \mathcal{D}_\mu\Phi^\dag\mathcal{D}^\mu\Phi + \lambda(\Phi^\dag\Phi - v^2)^2 + g^2\chi^2\Phi^\dag\Phi + \frac{1}{2}{\rm Tr}\lbrace F_{\mn}F^{\mn}\rbrace\,,
\end{equation}
with $g,\lambda$ dimensionless couplings, $F_{\mn}$ the non-abelian field strength and $\mathcal{D}_\mu = \partial_\mu - ieA_\mu$ the gauge derivative, coupling $\Phi$ with the gauge fields $A_\mu$ with strength $e$.

A realistic scenario of Reheating should account for (almost) all the matter of the Universe. Therefore, it should really incorporate a complete gauge theory with all kind of fermions and bosons, since our actual understanding of particle physics relies on the Standard Model (SM), based on the $SU(3)\times SU(2)\times U(1)$ group involving scalar, spinor and gauge fields. Of course, the SM is known to be incomplete and there are several extensions of which one can think about. But whatever new ingredients might be added, this does not change the fact that SM particles had to be produced during Reheating, since the posterior evolution of the Universe cannot be understood if those particles were not already present in an early epoch. Let's then discuss a potential way to realize Reheating within the SM.

\section{The Standard Model Higgs Non-Minimally Coupled to Gravity}\label{section:SM Higgs Non Minimally Coupled}

Supposse we consider the SM Higgs field non-minimally coupled to gravity, such that the Universe is described by 
\begin{equation}\label{SHG}
 S = S_{\rm SM} + S_{HG} = S_{\rm SM} + \int d^4x \sqrt{-g} \{
    \frac{1}{2}M_P^2R+\xi \Phi^\dagger \Phi R \}\;,
\end{equation}
where $S_{\rm SM}$ is the SM action, $M_p \approx 2.43\cdot10^{18}$ GeV is the reduced Planck mass, $R$ is the Ricci scalar, and $\xi$ represents the strength of the Higgs-Gravity (dimensionless) coupling. In the unitary gauge, the Higgs field can be represented as $\Phi = h/\sqrt{2}$. The Higgs-gravity sector in the Jordan (J) frame then takes the form
\begin{eqnarray}\label{higgslagrangJ}
\int d^4x \sqrt{-g} \Big[f(h)R - 
 \frac{1}{2}g^{\mu\nu}\partial_\mu \,h\partial_\nu h - U(h) \Big]\hV\\ 
\label{eq:potentialJ}
f(h)=(M_P^2+\xi h^2)/2\,,\hV U(h)=\frac{\lambda}{4}\left(h^2-v^2\right)^2\,,
\end{eqnarray} 
where $U(h)$ is the usual ElectroWeak Spontaneous Symmetry Breaking (EW SSB) potential of the SM Higgs, with vacuum expectation value (\textit{vev}) $v=246$ GeV and self-coupling $\lambda \sim \mathcal{O}(0.1)$.

As first discussed in~\cite{Bezrukov:2007}, if we require this model to be responsible for inflation, the parameters $\xi$ and $\lambda$ must be related as $\xi\simeq 10^5\sqrt{\lambda}$. The reason is that while the amplitude of Cosmic Microwave Background (CMB) anisotropies fixes the self-coupling of a quartic potential to be $\lambda\sim\mathcal{O}(10^{-13})$~\cite{Linde90}, the addition of the non-minimal coupling changes that condition to $\lambda \sim 10^{-10}\xi^2$~\cite{Bezrukov:2007}. Since we are thinking of the SM Higgs, for which $\lambda \sim \mathcal{O}(0.1)$, we are then forced to require $\xi \sim \mathcal{O}(10^{4})$. For more details see~\cite{Bezrukov:2007}, or the more recent discussions~\cite{HiggsDrivenInflation}. For a critical viewpoint see~\cite{criticsHiggsDrivenInflation}. Here we just want to stress that a large non-minimal coupling is a fundamental ingredient if the SM Higgs is to be responsible for inflation. Thus, assuming that~(\ref{SHG}) correctly describes the early Universe at the required energy scale for inflation (see next), we will only focus here on the details of Reheating~\cite{BezrukovReheating},\cite{Figueroa4}, just after the inflationary period.

Analyzing Reheating with a (significant) non-minimal coupling in the Jordan frame can be challenging, so let us first perform a conformal transformation to the metric, $g_{\mu\nu} \rightarrow \tilde{g}_{\mu\nu}=\Omega^2 g_{\mu\nu}$. By imposing the condition $f(h)/\Omega^2\equiv M_P^2/2$, we can this way obtain the usual Einstein-Hilbert gravitational term. From such a condition one then finds the relation
\begin{equation}
\label{relwithomega} 
 \Omega^2(h) = 1 + \frac{\xi h^2}{M_P^2}\,.
\end{equation}
The conformal transformation leads to a non-minimal kinetic term for the Higgs
field~\cite{Figueroa4}, which nevertheless can be reduced to a canonical one by redefining the Higgs as
\begin{equation} \label{relchih}
  \frac{d\chi}{dh}=\sqrt{\frac{\Omega^2+6\xi^2h^2/M_P^2}{\Omega^4}}=
  \sqrt{{1+\xi(1+6\xi)h^2/M_P^2 \over (1+\xi h^2/M_P^2)^2}}\,.
\end{equation}
Thus, considering the transformations~(\ref{relwithomega}) and (\ref{relchih}), the total action (ignoring the gauge interactions) in the conformally transformed frame -~{the Einstein frame}~-, is simply
\begin{eqnarray}
  \label{higgsLagrangianE3}
     \int d^4x\sqrt{- {g}} \, \Bigg[ \frac{M_P^2}{2} {R}
    - {1\over2}  g^{\mu\nu}\partial_\mu \chi\,\partial_\nu \chi - V(\chi)  \Bigg]\,,\\
 \label{eq:potentialE}
   V(\chi) \equiv \frac{U(h(\chi))}{\Omega^4(\chi)}\,,\hI\hVII
\end{eqnarray}
with $U(h)$ given by~(\ref{eq:potentialJ}). To find the explicit form of the potential~(\ref{eq:potentialE}) in terms of the new variable $\chi$, we must find $h$ as a function of $\chi$, by integrating Eq. (\ref{relchih}). In our case of a large coupling, $\xi\gg 1$, it can be shown~\cite{Figueroa4} that 
\begin{equation}\label{eq:relchih2}
 \Omega^2(h) \approx e^{\alpha\kappa\chi}\;,
\end{equation}
with $\alpha=\sqrt{2/3}$ and $\kappa=1/M_P$. From~({\ref{eq:relchih2}), one finds that the Higgs potential~(\ref{eq:potentialE}) in the Einstein frame looks like \begin{equation}
\label{eq:potentialE2}
V(\chi) \approx \frac{\lambda M_P^4}{4\xi^2}\left(1-e^{-\alpha\vert\chi\vert/M_{\rm p}}\right)^2\;,
\end{equation}
where we used the fact that $v \ll M_p$. During Inflation $\chi \gg M_p/\alpha$ and therefore the inflationary energy density is
\begin{equation}
 V_{\rm inf} \approx \frac{\lambda M_P^4}{4\xi^2} \sim M_{\rm GUT}^4,
\end{equation}
where $M_{\rm GUT} = 10^{16}$ GeV is the GUT scale. Thus, we already know the initial conditions for Reheating in this model. The Universe is filled with a homogeneous condensate of the Higgs field, whose energy density is around the GUT scale. During Reheating, the Higgs condensate will oscillate around the bottom of its potential. Therefore, in order to study the production of particles in the presence of a dynamic Higgs condensate, we first have to derive the couplings between the Higgs and the rest of the SM particles, in the Einstein frame.

In the Standard Model, the masses of the gauge bosons and of the fermions are given by the Higgs mechanism, after the Higgs acquires a constant \textit{vev}. In our case the Higgs field will evolve in time, $h = h(\chi(t))$, so the effective masses of the fermions and of the gauge bosons will then be changing in time as well, like
\begin{equation}\label{massesWZ}
m_W = m_Z\,{\cos\theta_W} = \frac{1}{2}g_2h(\chi(t))\,,\hV m_{f} = \frac{1}{2}y_f\,h(\chi(t))\,,
\end{equation}
where $\theta_W$ is the Weinberg angle $\theta_W=\tan^{-1}(g_1/g_2)$, and $y_f$, $g_1$ and $g_2$ are the Yukawa and the $U(1)_Y$ and $SU(2)_L$ couplings, respectively. The pieces of the SM action of interest for us are: the Spontaneous Symmetry Breaking sector, responsible for the $W$ and $Z$ gauge bosons masses, the Charged and Neutral Currents, coupling the SM fermions to gauge bosons through the $J_\mu^\pm$, $J^Z_\mu$ currents, and the Yukawa sector, coupling the SM fermions with the Higgs,
\begin{eqnarray}\label{SSBJ}
S_{SSB} = \int d^4x \sqrt{-g} \Big\{ m_W^2 W_{\mu}^+ W^{\mu -}+\frac{1}{2} m_Z^2 Z_\mu Z^\mu  \Big\},\hI\hV \\
\label{F}
\hI\hV S_{CC} + S_{NC} = \int d^4x \sqrt{-g} \left\{\frac{g_2}{\sqrt{2}}W_\mu^+J^-_\mu +\frac{g_2}{\sqrt{2}}W_\mu^-J^+_\mu+ \frac{g_2}{\cos\theta_W} Z_\mu J_Z^\mu\right\},\\
\label{yukawaJ}
S_{Y} = \int d^4x \sqrt{-g} \Big\{ m_d {\bar\psi_{d}}{\psi_{d}} + m_u { \bar\psi_{u}}{ \psi_u}\Big\},\hI\hI
\end{eqnarray}
By redefining the fields and masses with a specific conformal weight as
\begin{eqnarray}\label{eq:FieldsRedef}
\tilde{W}_\mu^{\pm}\equiv\frac{{W}_\mu^{\pm}}{\Omega}\;,\hI \tilde{Z}_\mu\equiv\frac{{Z}_\mu}{\Omega}\; \hI
\tilde{\psi}_{d}\equiv\frac{\psi_{d}}{\Omega^{3/2}}\;, \hI \tilde{\psi}_{u}\equiv\frac{\psi_{u}}{\Omega^{3/2}}\hI\hI \\
\label{eq:MassRedef}
\tilde{m}^2_W = \tilde{m}^2_Z\,\cos^2\theta_W = \frac{m^2_W}{\Omega^2}=\frac{g^2_2 M_P^2(1-e^{-\alpha\kappa\vert\chi\vert})}{4\xi}
\;, \hI \tilde{m}_f \equiv \frac{m_f}{\Omega} = \frac{y_f M_P}{\sqrt{2\xi}}\left(1-e^{-\alpha\kappa\vert\chi\vert}\right)^{1/2}
\;,
\end{eqnarray}
the three pieces of the SM action~(\ref{SSBJ}) preserve their functional form in the Einstein frame. In other words, the Lagrangian describing the gauge interactions between the SM fields, can be written in the Einstein frame identically as Lagrangian~(\ref{SSBJ}) in the Jordan frame, as long as the redefinitions~(\ref{eq:FieldsRedef}),(\ref{eq:MassRedef}) are considered. A key point here is that, not only the form of the interaction terms is known from the gauge principle, but also the strength of the couplings is known from the high energy particle physics experiments. Most of the work in the literature on Reheating had only been focused on models encoding the different mechanisms that could play a role in the process, with the strength of the couplings set essentially by hand. However, in the model under discussion, the relative importance of each mechanism of particle creation can be exactly asserted, since for the first time we are studying Reheating in a scenario in which the field content, the form of the interactions and the strength of the couplings, are all fixed. From this point of view, we can say that Reheating in this Higgs-driven inflation scenario, is a realistic scenario of Reheating.

\section{Reheating: Transferring the energy to the SM particles}\label{sec:Reheating}

In order to analyze Reheating, let us first find the dynamics of the Higgs condensate. Soon after inflation ends, the effective Higgs potential~(\ref{eq:potentialE2}) can be approximated by a simple quadratic potential around the minimum,
\begin{eqnarray}\label{potentialEapprox}
V(\chi) = {1\over2}M^2\chi^2 + \Delta V(\chi)\,,\hI
M = \sqrt{\frac{\lambda}{3}}\frac{M_P}{\xi} \sim 10^{-5}M_p\,,\hV
\end{eqnarray}
where $\Delta V$ are some corrections which become negligible during Reheating (see next). Thus, at the end of Inflation, the Higgs condensate is formed by Higgs quanta of momentum ${\bf k} = {\bf 0}$ (corresponding to a homogeneous field) and mass $M$. This might seem surprising to the reader familiar with the SM ElectroWeak (EW) Spontaneous Symmetry Breaking (SSB) process, since there the mass of the Higgs excitations around the true vacuum, is known to be $\sqrt{\lambda}v/\sqrt{2} \ll M$. However, let us remind the that the EW SSB process, as indicated by its name, takes place at the EW scale $E_{\rm EW} \sim 10^2$ GeV, whereas here the energy scale is $V_{\rm inf}^{1/4} \sim E_{\rm GUT} \sim 10^{16}$ GeV. We are considering a very large field approximation of the SM Higgs SSB potential, $|\chi| \gg v$, as induced through the non-minimal coupling to gravity and the requirement of inflation. At the inflationary scale, the system does not see the static Higgs $vev$ $v$, and only the dynamical amplitude~(\ref{physsolution}) of $\chi$ will matter, being $\chi \gg v$.

Considering that the Universe expands with scale factor ${a\propto t^{q}}$, then the equation of the homogeneous Higgs is
\begin{eqnarray}\label{eqNonLinearHiggs}
\ddot\chi + 3H\dot\chi + V'(\chi) = 0\hV\Rightarrow\hV t^2\ddot\chi + 3qt\dot\chi + t^2M^2[1+\delta M^2(\chi)]\chi = 0\,,
\end{eqnarray}
with $\delta M^2(\chi)$ describing the non-linear corrections due to the Higgs'~self-interactions. In particular, expanding~(\ref{eq:potentialE2}) around the minimum, $\delta M^2 = -\beta|\chi| + \zeta\chi^2 + {\cal O}(\chi^3)\,,$ with $\beta=\lambda M_P/\sqrt{6}\xi^2$ and $\gamma=7\lambda/27\xi^2$. Let us then consider than those corrections are indeed negligible from the beginning of Reheating, i.e. $|\delta M^2(\chi)| \ll 1$. Then we'll see if this can be justified \textit{a posteriori}. Neglecting such terms, the solution to~(\ref{eqNonLinearHiggs}) is $\chi(t) \propto (Mt)^{-\nu}J_{\nu}(Mt)$, with $J_{\nu}(x)$ the Bessel function of order $\nu = (3q-1)/2$. Making use of the large argument expansion ($Mt\gg1$) of Bessel functions, we then find that the amplitude of the Higgs condensate evolves as
\begin{equation}\label{physsolution}
\chi(t) \approx X(t)\sin(Mt)\,,\hI \,\, X(t) \propto (Mt)^{-\frac{3q}{2}}
\end{equation}
Thus, the resulting dynamics of the homogeneous Higgs condensate correspond to an oscillatory field with (angular) frequency $M$ and decaying amplitude $X(t)$. The energy and pressure densities associated to the $\chi$ field, as described by~(\ref{physsolution}), can be obtained after averaging over several oscillations, as
\begin{eqnarray}\label{energyInflaton}
2\rho_\chi \approx \left\langle \dot\chi^2+ M^2\chi^2 \right\rangle &\approx& M^2X^2 [\left\langle\cos^2(Mt)\right\rangle + \left\langle\sin^2(Mt)\right\rangle] = M^2X^2\;, \\
2p_\chi \approx\left\langle\dot\chi^2 - M^2\chi^2\right\rangle &\approx& M^2X^2[\left\langle\cos^2(Mt)\right\rangle -\left\langle\sin^2(Mt)\right\rangle] =  0\;, 
\end{eqnarray}
Since the averaged pressure is negligible, then $q \equiv d\log a(t)/d\log{\rm t} $ is forced to be $2/3$, as if a matter-dominated background would dictate the expansion rate. Using this, the physical solution can then finally be expressed as
\begin{eqnarray}\label{chit}
\chi(t) \approx \frac{\chi_{\rm e}}{2\pi N}\sin(2 \pi N)=\frac{\chi_{\rm e}}{j\pi}\sin(\pi j)\equiv X(j) \sin(\pi j)\;,
\end{eqnarray}
where $j = (Mt)/\pi$ is the number of times the Higgs crosses around zero and $N$ is the number of oscillations $N = j/2$. Using now the the (covariant) energy conservation equation, $\dot\rho_\chi = - 3(\dot a/a)\rho_\chi$, we can then find that $\chi_{\rm e} = \sqrt{8/3}\,M_p$. Introducing now~(\ref{chit}) into $\delta M^2$, we find that even just after the first oscillation, $|\delta M^2| \ll 0.122$. Thus, from the very end of inflation, the Higgs effective potential tends very rapidly to that of a (damped) harmonic oscillator, which justifies self-consistently \textit{a posteriori} the approximation $\vert \delta M^2\vert \ll 1$ used in the derivation of~(\ref{chit}).

Let us now move into the details of the transfer of energy from the dynamical Higgs condensate to the SM particles. We will first look at the (perturbative) decay of the Higgs, and realize that before a single Higgs particle decays, more interesting (non-pertubative) phenomena take place. We will then show that, as in the standard case of Reheating after chaotic inflation, particles are created when the Higgs condensate oscillates around the minimum of its potential. However, the produced particles significantly decay into other species during each oscillation of the Higgs and, as a result, a new phenomenological interplay of perturbative and non-perturbative effects has to be taken into account.

\subsection{Frustration of the Higgs Perturbative Decay}\label{subsec:Frustration}

A natural Reheating mechanism, would be a perturbative decay process of the Higgs quanta into the SM particles, right after Inflation ended. In order to have a perturbative decay of a Higgs particle two conditions must be fulfilled:

1) The Higgs decay rate $\Gamma\sim \frac{g^2}{8\pi}M$ has to be greater than the rate of expansion $H^2=\frac{\rho_\chi}{3M_P^2} \approx \frac{1}{6}(\frac{M}{M_P})^2\left(\frac{\chi_{\rm end}}{\pi j}\right)^2$.

2) There should be enough phase-space in the final states for the Higgs field to decay, i.e. $M>2m_{f},m_{A}$.
\vskip.15cm
From condition 2), we can deduce~\cite{Figueroa4} that the Higgs condensate should need to oscillate $\sim 10^{6}$ times, before it is (statistically) allowed to decay into gauge bosons. And the same applies to top quarks. In the case of the decay into bottom and charm quarks, this channel is opened only after a few oscillations of the Higgs, while for the rest of quarks and leptons, the decay-channel has sufficient phase space from the very end of inflation. In general, the smaller the Yukawa coupling of a given fermion species to the Higgs, the less oscillations the Higgs has to perform before there is enough phase-space to decay into such fermion species. However, the smallness of the Yukawa coupling implies also a smaller decay rate. Unfortunately, for every species of the SM,  when there is phase-space for the Higgs to decay into such species, the decay rate does not catch up with the expansion rate. Vice-versa, if the decay rate of a given species overtakes the expansion rate, there is no phase-space for the Higgs decay to happen. Therefore, during a large number of oscillations, the Higgs is not allowed to decay perturbatively into any of the Standard Model fields. Fortunately, before any of such decay channels is opened, many other interesting (non-perturbative effects) start taking place. 

\subsection{Non-Perturbative Production of Gauge Fields}\label{NoAdiabaticity}

Consider for instance the interaction of the Higgs field with the Z gauge bosons~(\ref{SSBJ},\ref{eq:FieldsRedef},\ref{eq:MassRedef}). Around the minimum of the Higgs potential, such interaction can be approximated as a tri-linear term $|\chi|Z_\mu Z^\mu$, since the effective mass~(\ref{eq:MassRedef}) of the Z gauge boson can be approximated for a small Higgs field amplitude as
\begin{equation}\label{eq:massZ}
\tilde{m}^2_Z\simeq \frac{\alpha g^2_2 M_P}{4\xi \cos^2\theta_W} \vert\chi\vert\;,
\end{equation}
For the main part of an oscillation of $\chi$, $\tilde{m} \gg M$ and, as a result, the Z gauge boson will oscillate many times during each oscillation of the Higgs. Of course, the same applies to the W bosons. Moreover, from~(\ref{eq:massZ}) and~(\ref{physsolution}), we can see that during most of the time of Higgs oscillation, the effective masses of the intermediate bosons are changing adiabatically, verifying the condition $\vert \dot{\tilde{m}}\vert \ll \tilde{m}^2$. However, for values of $\chi$ close to zero, the adiabatic condition is violated, and the passage of $\chi$ through the minimum of the potential can be interpreted as particle production~\cite{Preheating}. In particular, using the fact that the Higgs velocity around zero can be approximated as $\dot \chi(j) \approx MX(j)$, the violation of adiabaticity can be found in correspondence with a region $|\chi| < \chi_a$~\cite{Figueroa4}, with
\begin{equation}\label{chia}
\chi_a=\left(\frac{\xi \vert\dot\chi(t)\vert^2}{\alpha g^2 M_P}\right)^{1/3}= \,\,\left(\frac{\lambda\pi}{4 g^2\xi}\right)^{1\over3}j^{1/3}\,X(j)\,, 
\end{equation}
where (from now on, unless otherwise stated), $g = g_2,\,\, g_2/\cos\theta_W$ for the $W$ or $Z$ bosons respectively\footnote{All the couplings are renormalized at the inflationary energy scale, $V_{\rm Inf}^{1/4} \sim M_{\rm GUT}$, so here $g^2_1 \approx g^2_2 \approx 0.30$ and $\sin^2\theta_W = \cos^2\theta_W \approx 0.7$.}. Only outside this region, $|\chi| > \chi_a$, the notion of particle makes sense and an adiabatic invariants can be defined. The previous regions are indeed very narrow compared to the amplitude of the oscillating Higgs, $\chi_a \sim 10^{-2}j^{1/3}X(j)$. Therefore, the particle production taking place in such a narrow field region, happens within a very short interval of time as compared to the Higgs period of oscillation $T=2\pi/M$, which can be estimated~\cite{Figueroa4} as $\Delta t_a(j)\sim {2\chi_a}/{\vert\dot\chi\vert}\sim 10^{-2}\,j^{1/3}M^{-1} \ll T$. Note that, independently of the species, $W$ or $Z$ bosons, many oscillations ($N \sim 10^3$) will pass before the fraction of time spent in the non-adiabatic zone will increase from a 1\% to a 10\%, as compared with the period $T$. 

We will now discuss the non-perturbative creation of particles in the non-adiabatic region. Expanding Eq.~(\ref{chit}) around the $j$-th zero at time $Mt_j = \pi j$, the evolution equation of the bosonic fluctuations can be approximated (see~\cite{Figueroa4} for more details) as a Schr\"odinger-like equation of the type
\begin{eqnarray}\label{schrodinger}
-W_k'' - \frac{q_W}{j}|\tau|W_k = K^2 W_k\;, \hspace{15mm} -Z_k'' - \frac{q_Z}{j}|\tau|Z_k = K^2 Z_k\;, 
\end{eqnarray}
where primes denote derivatives with respect to the rescaled time $\tau = Mt$, $K \equiv \frac{k}{aM}$, and
\begin{eqnarray}\label{qfactor}
q_W = \cos\theta_W^2 q_Z= \frac{g_2^2\alpha\kappa\chi_{\rm e} }{4\pi\xi}\left(\frac{M_p}{M}\right)^2 = \frac{g_2^2\xi\,}{\pi\lambda}\;, 
\end{eqnarray}
are the usual resonance parameters~\cite{Preheating}. Therefore, each time the Higgs crosses zero, we can formally interpret such event as analogous to the quantum mechanical scattering of a particle crossing an inverted triangular potential. In particular, let $T$ and $R = 1- T$ be the transmission and reflection probabilities for a single scattering in this triangular barrier. The number of particles just after the $j$-th scattering, $n_{k}(j^+)$, in terms of the previous number of particles $n_{k}(j^-)$ just before that scattering, can be written as~\cite{Preheating}
\begin{eqnarray}\label{occupnum}
n_{k}(j^+) = C(x_j) + [2C(x_j) + 1]n_{k}(j^-) +2\cos\theta_j {\sqrt{C(x_j)\left[C(x_j)+1\right]}}\sqrt{n_{k}(j^-)\left[n_{k}(j^-)+1\right]}\,,
\end{eqnarray}
where $\theta_j$ are some accumulated phases at each scattering (which we will discuss later), and $C(x_j)$ is related to the transmission probability $T_{k}(j)$ for the $j$-th scattering as~\cite{GalindoPascualbook}
\begin{eqnarray}\label{Transmision}
C(x_j) \equiv T^{-1}_{k}(j)-1 = \pi^2\left[{\rm{Ai}}\left(-x_j^2\right){\rm{Ai}}'\left(-x_j^2\right) + {\rm{Bi}}\left(-x_j^2\right){\rm{Bi}}'\left(-x_j^2\right)\right]^2\,,\hV 
x_{j} \equiv \frac{j^{1/3}k}{Mq^{1/3}a_j}\,,
\end{eqnarray}
with $a_j$ the scale factor at $t_j = \pi j/M$ and $\rm{Ai}(z),\rm{Bi}(z)$ the Airy functions. Normalizing the scale factor at the first zero crossing as $a_1 = 1$, we can then simply write the evolution of the scale factor as $a_j = j^{2/3}$. Thus, the behavior of $x_j$ with the number of zero crossings goes as $\propto j^{-1/3}$ and from here, the natural range for the momenta of the problem, $k_*{(j)}$, can be found in terms of the $q$ resonant parameters~(\ref{qfactor}), as
\begin{equation}\label{typicalMoment}
x_j = 1 \hV \Rightarrow \hV k_*{(j)} \equiv 
q^{1/3}\,j^{1/3}\,M.
\end{equation}

Considering the situation in which $n_{k}(j^-) \ll 1$, certainly true in the first scatterings, then
\begin{eqnarray}\label{deltaN}
\Delta n_{k}(j^+) \approx C(x_j) \equiv T^{-1}_{k}(j)-1\,,
\end{eqnarray} 
where we have retained only the first term of Eq.~(\ref{occupnum}). This corresponds to the spontaneous particle creation of $W$ and $Z$ bosons each time the Higgs crosses around zero. In particular, the total number of produced particles of a given species (and polarization), just after exiting the non-adiabatic region around the $j$-th zero-crossing, can be obtained as
\begin{equation}\label{spontaneousCreation}
\Delta n(j^+) =\frac{1}{2\pi^2\,a_j^3}\int_0^\infty dk\,k^2\,C(x_j) =\frac{q}{2j}\,\mathcal{I}\,M^3\,,
\end{equation}
with $\mathcal{I} = {\pi}^{-1}\int_0^\infty C(x)x^2dx \approx 0.0046$ and $q$ the resonant parameters given by Eq.~(\ref{qfactor}). Thus, the only difference between the number of $W$ and $Z$ bosons produced is simply encoded in the different resonance parameter, $q_W \propto g_2^2$ and $q_Z \propto g_2^2/\cos^2\theta_W$.

As pointed out first in~\cite{Preheating}, one would expect that during the first oscillations of the inflaton (the Higgs in our case), the first term of~(\ref{occupnum}) would dominate over the others, and the particle production should be driven by spontaneous creation at the bottom of the potential, as described by~(\ref{spontaneousCreation}). Eventually, after some oscillations, a significant number of particles would have been created and, consequently, the terms proportional to $n_k$ in~(\ref{occupnum}) should dominate. Those terms, as opposed to~(\ref{deltaN}), induce a stimulated growth of particles in an explosive manner, known as {\tt parametric resonance}. The accumulation of particles is thus expected to enhance the rate of production, entering into a regime of exponential growth. At least, that is the usual picture expected for Reheating~\cite{Preheating} in a scenario in which the inflaton oscillates around its potential. However, in our scenario under discussion, a new phenomenon is going to occur. 

\subsection{Perturbative Decays of Gauge Bosons}\label{GaugeBosonsPerturbativeDecays}

In the Reheating scenario under analysis we have access to the form and strength of the couplings between all particles. Therefore, we can compute the total decay widths of the $W^{\pm}$ and $Z$ bosons into any pair of fermions. As opposed to the conventional result found in particle physics books, now the gauge bosons' masses are time-dependent functions of the dynamical $vev$ of the Higgs and, as a consequence, the decay widths will vary in time during Reheating. The functional form of the decay rate in the Einstein frame will preserve the same form as in the Jordan frame, being only changed through the conformally transformed masses, see Eq.~(\ref{eq:MassRedef}). After some calculus, one finds
\begin{eqnarray}
\label{widthWE}
\Gamma_{W^\pm}^E = \frac{3g_2^2 \tilde{m}_W}{16\pi} = \frac{3g_2^3M_p}{32\pi\xi^{1/2}}\left(1-e^{-\alpha\kappa|\chi|}\right)^{1/2}\,,\hI \Gamma^E_{Z} = \frac{2{\rm Lips}}{3\cos^3\theta_W}\,\Gamma_{W^\pm}^E
\end{eqnarray} 
where ${\rm Lips}\equiv\frac{7}{4}-\frac{11}{3}\sin^2\theta_W+\frac{49}{9}\sin^4\theta_W$ denotes the \textit{Lorentz invariant phase-space} factors. Note that after the passage of $\chi$ through the bottom of the potential, the gauge bosons' masses grow as the Higgs climbs up the potential and, as a result, the decay width also increases significantly. The $W$ and $Z$ bosons then tend to decay into fermions in a time $\Delta t\sim 1/\langle\Gamma^E_{W,Z}\rangle_j$, where $\langle \cdot \rangle_j$ represents a time average between the $j$- and the $(j+1)$-th Higgs zero crossings. The typical time of decay turns out to be $\Delta t \simeq \epsilon\, j^{1/2}\, M^{-1}$, with $\epsilon = 0.64$ or $1.55$ for $Z$ and $W$ bosons, respectively. Therefore during the first oscillations of the Higgs, the non-perturbatively produced gauge bosons significantly decay within a semiperiod $T/2 = \pi M^{-1}$. As the amplitude of the Higgs field decreases with time, the decay width of the gauge bosons~(\ref{widthWE}) also decreases and becomes less and less significant. However, as we will explain next, simply the fact that bosons decay during the first oscillations of the Higgs, will have important consequences.

\subsection{Combined Preheating: Mixing Perturbative and Non-Perturbative effects}\label{subsec:CombinedPreheating}

The total number density of gauge bosons $n(j^+)$ present just after the $j$-th crossing will decay exponentially fast until the next crossing, due to the perturbative decay into fermions. Therefore the total number density just previous to the $(j+1)$-th zero crossing, $n((j+1)^-)$, is given by
\begin{eqnarray}\label{eq:decay}
n((j+1)^-) = n(j^+)e^{-\int_{t_j}^{t_{j+1}} \Gamma dt} = n(j^+)e^{-\langle\Gamma\rangle_j\frac{T}{2}}\;.
\end{eqnarray} 
On the other hand, in the large occupation limit $n_k \gg 1$, we can neglect the first term in Eq.~(\ref{occupnum}). We then find that the spectral number densities of the gauge bosons just after and previous to the $j$-th scattering, $n_k(j^+)$ and $n_k(j^-)$, respectively, become proportional to each other as
\begin{eqnarray}\label{expGrowth}
n_k(j^+) \approx \left( (2C(x_j)+1) - 2\cos\theta_j\sqrt{C(x_j)(C(x_j)+1)}\right)n_k(j^-) \equiv n_k{(j^-)}e^{2\pi\mu_k(j)}\,,
\end{eqnarray}
where $C(x_j)$ was defined in~(\ref{Transmision}) and $\mu_k(j)$ is the Floquet index. The $\{\theta_j\}$ are some accumulated phases at the $j$-th scattering, which can indeed play a very important role, since they can enhance ($\cos\theta_j < 0$) or decrease ($\cos\theta_j > 0$) the effect of particle production at each crossing. In particular, we can estimate~\cite{Figueroa4} them as
\begin{eqnarray}\label{phases}
\Delta\theta_j = \int_{t_j}^{t_{j+1}}dt\sqrt{k^2+{\tilde m}^2} \approx
\frac{g\pi\sqrt{3\xi}}{2\sqrt{\lambda}}F(j) \sim \mathcal{O}(10^{2})j^{-1/2}\,,\\
\label{Fn}
F(j) \equiv  \frac{1}{\pi}\int_{j\pi}^{(j+1)\pi} dx\left(1-e^{-\alpha\kappa|\chi(x)|}\right)^{1/2} \approx 0.43\,j^{-1/2}\,,
\end{eqnarray}
When $\Delta\theta_j \equiv \theta_{j+1} - \theta_{j} \gg \pi$, the effect of resonance will be chaotic, such that then the phases essentially are random at each scattering. From~(\ref{phases}), we conclude that such stochastic behavior occurs for the first $\mathcal{O}(10^{3})$ oscillations. Thus, for the first thousand oscillations the successive scatterings are incoherent, what allow us to define for each scattering, an average Floquet index obtained as $2\pi\bar{\mu}_k = \int_0^{2\pi} \mu_k(\theta)\,d\theta$, see~\cite{Figueroa4} for more details.

On the other hand, the decay widths~(\ref{widthWE}) of the gauge bosons, averaged over the Higgs oscillations, are given by 
\begin{eqnarray}
\left\langle\Gamma_{Z\rightarrow all}\right\rangle_j = \left(\frac{g_2 }{\cos\theta_W}\right)^3\frac{M_P\,{\rm Lips}}{16\pi\sqrt\xi}\,F(j) \equiv \frac{2\gamma_Z}{T} F(j)\,,\hI 
\left\langle\Gamma_{W\rightarrow all}\right\rangle_j = \frac{3\cos^3\theta_W}{2{\rm Lips}}\left\langle\Gamma_{Z\rightarrow all}\right\rangle_j\equiv \frac{2\gamma_W}{T} F(j)\;,\\
\label{smallGamma}
\gamma_Z = \left(\frac{g_2}{\cos\theta_W}\right)^{3}\frac{\sqrt{3}\xi^{1/2}}{16\lambda^{1/2}}\,{\rm Lips} \approx 14.23\lambda^{-\frac{1}{4}}\,, \hI \gamma_W \equiv \frac{3\cos^3\theta_W}{2{\rm Lips}}\gamma_Z \approx  5.91\lambda^{-\frac{1}{4}}\hI\hI
\end{eqnarray}
where $F(j)$ is defined in~(\ref{Fn}) and the constants $\gamma_Z,\gamma_W$ are just numerical factors depending of the parameters of the model and the decaying species. The decay of the vector bosons occurs precisely between two successive Higgs zero-crossings. Thus, taking into account Eqs.~(\ref{expGrowth}) and~(\ref{eq:decay}), the number of gauge bosons just after the $(j+1)$-th scattering, in terms of the number just after the previous one, can then be expressed as
\begin{equation}\label{formulaCrecimientorecursivo}
n_{k}((j+1)^+) = n_{k}((j+1)^-)e^{2\pi\mu_{k}(j+1)} = n_{k}(j^+)e^{-\gamma\,F(j)}e^{2\pi\mu_{k}(j+1)}\,,
\end{equation}
This formula~(\ref{formulaCrecimientorecursivo}) clearly shows how the two effects, the non-perturbative resonance production (via $e^{2\pi\mu_k}$) and the perturbative decays (through $e^{-\gamma F(j)}$), are combined together. This combination indeed generates new phenomenology and to emphasize the difference from the usual parametric resonance, we will call it $Combined$ $Preheating$. Applying~(\ref{formulaCrecimientorecursivo}) recursively, we could estimate the occupation number (for each species and polarization) just after the $j$-th scattering, $n_k(j^+)$. That was indeed the procedure we adopted in~\cite{Figueroa4}. However here, to be more accurate, we will rather use the recursive iteration of~(\ref{eq:growth2}). The reason is that we obtained~(\ref{formulaCrecimientorecursivo}) by neglecting the first term of~(\ref{occupnum}), which is a reasonable approximation when parametric resonance takes place very soon after the end of inflation (as implicitly assumed in chaotic models). However, in the model we are discussing, the gauge bosons decay significantly. Thus, they are intially not sufficiently accumulated and, consequently, resonance is delayed. In particular, considering~(\ref{occupnum}) and~(\ref{eq:decay}) all together, one obtains the phase averaged relation
\begin{eqnarray}\label{eq:growth2}
\left(\frac{1}{2}+n_k((j+1)^+)\right) = (1+2{\rm C}(x_j))\left(\frac{1}{2}+n_k(j^+)\,e^{-\gamma F(j)}\right)\,,
\end{eqnarray}
where we simply dropped out the term involving $\cos\theta_j$, after averaging Eq.~(\ref{occupnum}) over the random phases $\{\theta_j\}$. We clearly see that the perturbative decays $e^{-\gamma F(i)}$ tend to decrease the rate of production of bosons, while the factors $(1+2C(x_j)) > 1$ [or equivalently the factors $e^{2\pi\mu_k}$ in~(\ref{formulaCrecimientorecursivo})] tend to (resonantly) enhance it, due to the accumulation of previously produced bosons. Initially, the perturbative decays prevent the resonance to be effective. However, after a certain number of oscillations, the resonant effect may will finally be developed, since the perturbative decays become less and less important as times goes by. That is precisely the novelty in Combined Preheating, a simultaneous competition between perturbative decays and non-perturbative production of particles, affecting each other recursively. Depending on the couplings, the energy density transferred into the fermionic species might dominate over the Higgs condensate before parametric resonance becomes relevant or, perhaps, the energy transferred into the fermions is negligible and the perturbative decays simply delay the development of parametric resonance. Since in this model we know the strength of the couplings, we will univocally determine which of the two previous options is the correct one. However, note that Combined Preheating might be a common ingredient (ignored so far) to many Reheating scenarios. As soon as the inflaton is coupled to some primary fields and the latter to some secondary fields, a competition between the non-perturbative production of the primary particles and a pertubative production of the secondary (from the decay of the primary ones) will take place. The rate at which energy is exchanged between the inflaton and the primary and secondary fields, could only be determined in each specific scenario if the involved couplings were known. However, the couplings of the inflaton to other matter fields are essentially free parameters in most models. Therefore, in the light of all this, we plan to explore in the future the space of parameters of other models different than the one under discussion, in order to determine the role that Combined Preheating could play. 

Coming back to our model, it can be shown~\cite{Figueroa4} that both $W$ and $Z$ bosons are non-relativistic, while their fermionic decay products are all ultra relativistic. Thus, the mean energy of the fermions (F) produced from the decay of the gauge bosons (B) between $t_j$ and $t_{j+1}$, can be estimated as
\begin{eqnarray}\label{fermionsEnergy}
E_{F({\rm B})}{(j)} \equiv \langle\,\,(k_F^2 + m_F^2)^{1/2}\, \rangle_j \approx \langle k_F \rangle_j \approx 0.5\langle \tilde m_B \rangle_j \approx 0.25 (g/\xi^{1/2})\,F(j)\,M_p\,,
\end{eqnarray}
Then, the number of fermions produced between the $j$-th and the $(j+1)$-th scatterings and their corresponding energy density, are simply given by
\begin{eqnarray}
\Delta n_{F}(j) = 2\times3\times\left(n_Z(j^+)(1-e^{-\gamma_ZF(j)}) + 2\,n_W(j^+)(1-e^{-\gamma_WF(j)})\right)\;,\hI\\
\label{deltaRhofer}
\Delta\rho_F{(j)} = 2\times3\left[(1-e^{-\gamma_Z F(j)})n_Z(j^+)E_{F({\rm Z})}(j) + 2(1-e^{-\gamma_WF(j)})n_W(j^+)E_{F({\rm W})}(j)\right]
\end{eqnarray}
where the factor $2\times3$ takes into account that each gauge boson can have one out of three polarizations and decay into two fermions, while the extra factor $2$ in front $n_W$, accounts both for the $W^+$ and $W^-$ decays. After $j$ crossings, the total energy density transferred into the relativistic fermions and into the non-relativistic bosons, can be expressed as
\begin{eqnarray}\label{rhoFer}
\rho_F{(j)} = \sum_{i=1}^{j} (i/j)^{8/3}\Delta\rho_F{(i)}\,,\hI\\
\label{rhoBos}
\rho_B(j) = 3\left(n_Z(j^+)\langle m_Z \rangle_{j} + 2n_W(j^+)\langle m_W \rangle_{j}\right)\,,
\end{eqnarray}

\begin{figure}
\centering
\includegraphics[width=5cm,height=8cm,angle= -90]{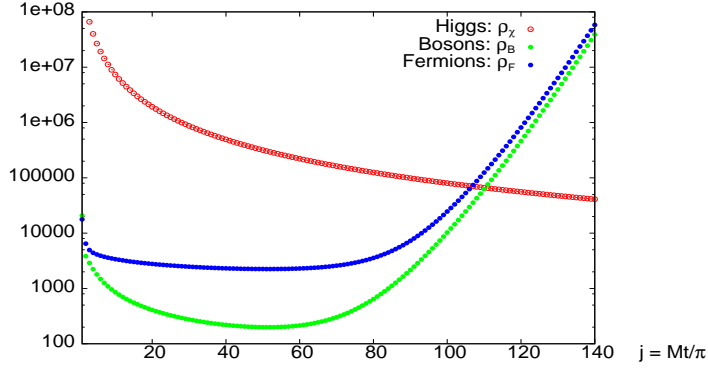}
\caption{Evolution of the energy density transferred into the gauge bosons and into the fermions as a function of $j$, for $\lambda = 0.2$ and $\xi = 44700\sqrt{\lambda}$. The decay of the homogeneous energy density of the Higgs is also shown. All densities are in units of $M^4$.}
 \label{EnergyEvolution}
\end{figure}

\noindent By iterating recursively~(\ref{eq:growth2}) we can find $n_Z(j^+)$ and $n_W(j^+)$ at each crossing, and plug them into~(\ref{deltaRhofer}),(\ref{rhoFer}) and~(\ref{rhoBos}). This way, we can follow the time evolution of the energy density of the Bosons and of the Fermions. The results are summarized in Fig.~(\ref{EnergyEvolution}). We find that the number of zero-crossings $j_{R}$ for which the perturbative decays stop blocking resonance, is $j_{R} \sim 70$ for the $W$ bosons and $j_R \sim 300$ for the $Z$ bosons. Parametric resonance thus becomes important much earlier for $W$'s than for $Z$'s since their decay rate~(\ref{widthWE}) differ in a factor $\gamma_Z/\gamma_W \approx 2.4$, such that many more $W$ bosons survive per half period than $Z$ bosons. Therefore, the \textit{Combined Preheating} of the $W$ bosons is much faster driven into the parametric-resonant like behavior, while the evolution of the $Z$ bosons is much more affected by the perturbative decays, completely preventing the development of parametric resonance. After several dozens of oscillations, the transfer of energy from the Higgs to the gauge bosons is completely dominated by the channel into the $W$ bosons, since they become fully resonant while the $Z$ bosons are still severely affected by their perturbative decay. 

We can estimate the time in which finally the energy of the inflaton would be transferred efficiently to the fermions or the bosons. Defining that moment, respectively, like $\varepsilon_F{(j_{\rm eff})} \equiv \rho_F/\rho_\chi \equiv 1$ and $\varepsilon_B{(j_{\rm eff})} \equiv \rho_B/\rho_\chi \equiv 1$, one obtains the numbers in Table~\ref{tab:a}. Note also that the number of oscillations $j_{\rm eff}$ required for an efficient transfer of energy, depends on the parameter $\lambda$, although the overall order of magnitude does not change appreciably.

Unfortunately, before reaching the stage in which $\epsilon_{F,B} \sim 1$, the backreaction of the produced gauge fields into the homogeneous Higgs condensate becomes significant. We don't have space here to give a detail account of this, so we will simply summarize our findings in~\cite{Figueroa4}. There we found that the effective oscillatory frequency of the Higgs, once the Higgs - Gauge fields interactions are restored in~(\ref{eqNonLinearHiggs}), is given by an expression of the type $\omega^2_{\rm osc} = M^2[1 + \mathcal{O}(g j^{3/2}/\xi^{3/2})(n_{W}(j)/M^3)]$. Thus, since resonance is initially blocked, the second term is very small and $\omega_{\rm osc} \approx M$, as we have implicitly assumed all the time. However, when parametric resonance becomes efficient, $n_W$ grows exponentially fast within few Higgs oscillations, and the term in $\omega_{\rm osc}^2$ proportional to $n_W$ eventually dominates. We can determine the number of Higgs crossings, $j_{\rm br}$, such that for $j>j_{\rm br}$, backreaction of the bosonic fields cannot be ignored anymore. The results, as a function of $\lambda$, are summarized in Table~\ref{tab:a}. We clearly see that backreaction seems to become important at a time slightly earlier than that at which we were expecting the Higgs to have transferred efficiently its energy to the bosons and fermions. This means that our analytical estimates of these transfers were biased, and a careful numerical study of the process is required beyond backreaction.

\begin{table}
\begin{tabular}{||c | c c c c c||}
\hline
    \tablehead{1}{c}{b}{$\lambda$}
  & \tablehead{1}{c}{b}{$0.2$}
  & \tablehead{1}{c}{b}{$0.4$}
  & \tablehead{1}{c}{b}{$0.6$}
  & \tablehead{1}{c}{b}{$0.8$}
  & \tablehead{1}{c}{b}{$1.0$}   \\
\hline
$j_{\rm eff}^{(F)}$ & 107 & 111 & 113 & 114 & 115\\
$j_{\rm eff}^{(B)}$ & 111 & 113 & 115 & 116 & 117\\ 
\hline
\end{tabular}
\label{tab:a}
\hI
\begin{tabular}{||c | c c c c c||}
\hline
    \tablehead{1}{c}{b}{$\lambda$}
  & \tablehead{1}{c}{b}{$0.2$}
  & \tablehead{1}{c}{b}{$0.4$}
  & \tablehead{1}{c}{b}{$0.6$}
  & \tablehead{1}{c}{b}{$0.8$}
  & \tablehead{1}{c}{b}{$1.0$}   \\
\hline
$j_{\rm br}^{(B)}$ & 107 & 110 & 112 & 113 & 114\\ 
\hline
\end{tabular}
\caption{Left: Number of Higgs crossings (depending on $\lambda$) for efficient Reheating. Right: Number of Higgs crossings (depending on $\lambda$) for efficient Reheating.}
\label{tab:a}
\end{table}

\section{Final Remarks}

We have studied Preheating after an inflationary stage driven by the Standard Model (SM) Higgs field non-minimally coupled to gravity. We have found that Preheating is driven by a complex process, which we called Combined Preheating, in which perturbative and non-perturbative effects occur simultaneously.

The Higgs field, initially an oscillating coherent condensate, produces non-perturbatively $W$ and $Z$ gauge fields. These decay very rapidly into fermions, what prevents gauge bosons to accumulate and, consequently, blocks the usual parametric resonance. The energy transferred into the fermionic species is, nevertheless, not enough to reheat the Universe, and resonant effects are eventually developed. Soon after resonance becomes effective, also backreaction from the gauge bosons into the Higgs condensate becomes relevant. We have determined the time evolution of the energy distribution among the remnant Higgs condensate and the non-thermal distribution of the SM fermions and gauge fields, until the moment in which backreaction becomes important. Beyond backreaction our approximations break down and numerical simulations and theoretical considerations beyond this work are required, in order to study the evolution of the system until thermalization.  

\begin{theacknowledgments}

I would like to thank Juan Garc\'ia-Bellido, Guy D. Moore and Javier Rubio for very stimulating discussions. I am also grateful to the organizers of the ``Invisible Universe International Conference", Paris 2009. I acknowledge support from a FPU contract (Ref.~AP2005-1092), from the Spanish Research Ministry (MICINN) under contract FPA2006-05807, and from the 6th Framework Marie Curie Network ``UniverseNet" under contract MRTN-CT-2006-035863.
\end{theacknowledgments}


\begin{thebibliography}{99}

\bibitem{Perturbative}
A.D. Linde, Phys. Lett. {\bf 108B}, 389 (1982) ; A.D. Dolgov and A.D. Linde, Phys. Lett. {\bf 116B}, 329 (1982); L.F. Abbott, E. Fahri and M. Wise, Phys. Lett. {\bf 117B}, 29 (1982).

\bibitem{Preheating}
  L.~Kofman, A.~D.~Linde and A.~A.~Starobinsky,
  Phys.\ Rev.\ Lett.\  {\bf 73}, 3195 (1994) ;  
  Phys.\ Rev.\  D {\bf 56}, 3258 (1997) ;  P.~B.~Greene, L.~Kofman, A.~D.~Linde and A.~A.~Starobinsky,
  Phys.\ Rev.\  D {\bf 56}, 6175 (1997).

\bibitem{Hybrid}
G.~N.~Felder, J.~Garcia-Bellido, P.~B.~Greene, L.~Kofman, A.~D.~Linde and I.~Tkachev,
  Phys.\ Rev.\ Lett.\  {\bf 87}, 011601 (2001) ; G.~N.~Felder, L.~Kofman and A.~D.~Linde,
  Phys.\ Rev.\  D {\bf 64}, 123517 (2001).

\bibitem{FermionicPreheating}
  P.~B.~Greene and L.~Kofman,
  Phys.\ Lett.\  B {\bf 448}, 6 (1999).

\bibitem{BellidoTonyMarga}
  A.~D.~Gil, J.~Garcia-Bellido, M.~Garcia-Perez and A.~Gonzalez-Arroyo,
  Phys.\ Rev.\  D {\bf 69}, 023504 (2004) ; Phys.\ Rev.\ Lett.\  {\bf 100}, 241301 (2008) ; JHEP {\bf 0807}, 043 (2008).

\bibitem{Bezrukov:2007}
F.~L.~Bezrukov and M.~Shaposhnikov,
  Phys.\ Lett.\  B {\bf 659}, 703 (2008).

\bibitem{Linde90}
A.~D.~Linde,
  ``Particle Physics and Inflationary Cosmology'', Chur, Switzerland: Harwood (1990)..

\bibitem{HiggsDrivenInflation}
A.~De Simone, M.~P.~Hertzberg and F.~Wilczek,
  Phys.\ Lett.\  B {\bf 678}, 1 (2009) ; F.~L.~Bezrukov, A.~Magnin and M.~Shaposhnikov,
  Phys.\ Lett.\  B {\bf 675}, 88 (2009) ; F.~Bezrukov and M.~Shaposhnikov,
  JHEP {\bf 0907}, 089 (2009) ; A.~O.~Barvinsky, A.~Y.~Kamenshchik, C.~Kiefer, A.~A.~Starobinsky and C.~Steinwachs, arXiv: 0904.1698 \& arXiv: 0910.1041

\bibitem{criticsHiggsDrivenInflation}
C.~P.~Burgess, H.~M.~Lee \& M.~Trott,
  JHEP {\bf 0909} (2009) 103 ; J.~L.~F.~Barbon \& J.~R.~Espinosa,
  PRD {\bf 79}, 081302 (2009).

\bibitem{BezrukovReheating}
F.~Bezrukov, D.~Gorbunov and M.~Shaposhnikov,
  JCAP {\bf 0906}, 029 (2009) .

\bibitem{Figueroa4}
J.~Garcia-Bellido, D.~G.~Figueroa and J.~Rubio,
  Phys.\ Rev.\  D {\bf 79}, 063531 (2009) (arXiv:0812.4624 [hep-ph]).

\bibitem{GalindoPascualbook} 
A. Galindo and P.~Pascual, ``Mec\'anica cu\'antica'', Ed. Alhambra Universidad, Madrid (1978).  

\end{thebibliography}
\end{document}